\documentclass[a4paper,fleqn,usenatbib]{mnras}
\usepackage[T1]{fontenc}
\usepackage{ae,aecompl}

\usepackage{graphicx}	
\usepackage{amsmath}	
\usepackage{amssymb}	

\title[21cm power spectra/global signal combinations]{Constraining cosmology and ionization history with combined $21\,\textrm{cm}$ power spectrum and global signal measurements}

\author[Liu \& Parsons]{
Adrian Liu,$^{1,\,2}$\thanks{Hubble Fellow}\thanks{E-mail: acliu@berkeley.edu}
Aaron R. Parsons$^{1,\,3}$
\\
$^{1}$Department of Astronomy, University of California Berkeley, Berkeley, CA 94720, USA\\
$^{2}$Berkeley Center for Cosmological Physics, University of California Berkeley, Berkeley, CA 94720, USA\\
$^{3}$Radio Astronomy Laboratory, University of California Berkeley, Berkeley, CA 94720, USA
}

\pubyear{2015}

\begin{document}
\label{firstpage}
\pagerange{\pageref{firstpage}--\pageref{lastpage}}
\maketitle

\begin{abstract}
Improvements in current instruments and the advent of next-generation instruments will soon push observational $21\,\textrm{cm}$ cosmology into a new era, with high significance measurements of both the power spectrum and the mean (``global") signal of the $21\,\textrm{cm}$ brightness temperature. In this paper we use the recently commenced Hydrogen Epoch of Reionization Array as a worked example to provide forecasts on astrophysical and cosmological parameter constraints. In doing so we improve upon previous forecasts in a number of ways. First, we provide updated forecasts using the latest best-fit cosmological parameters from the \emph{Planck} satellite, exploring the impact of different \emph{Planck} datasets on $21\,\textrm{cm}$ experiments. We also show that despite the exquisite constraints that other probes have placed on cosmological parameters, the remaining uncertainties are still large enough to have a non-negligible impact on upcoming $21\,\textrm{cm}$ data analyses. While this complicates high-precision constraints on reionization models, it provides an avenue for $21\,\textrm{cm}$ reionization measurements to constrain cosmology. We additionally forecast HERA's ability to measure the ionization history using a combination of power spectrum measurements and semi-analytic simulations. Finally, we consider ways in which $21\,\textrm{cm}$ global signal and power spectrum measurements can be combined, and propose a method by which power spectrum results can be used to train a compact parameterization of the global signal. This parameterization reduces the number of parameters needed to describe the global signal, increasing the likelihood of a high significance measurement.
\end{abstract}




\section{Introduction}

In coming years, experiments targeting the highly redshifted $21\,\textrm{cm}$ line have the potential to transform both astrophysics and cosmology. Large scale cosmological surveys of the $21\,\textrm{cm}$ brightness temperature probe the density, ionization state, and spin temperature of hydrogen atoms in the intergalactic medium (IGM), making them powerful \emph{direct} probes of the reionization epoch, when the first generation of galaxies systematically ionized the IGM \citep{hogan_and_rees1979,scott_and_rees1990,madau_et_al1997,tozzi_et_al2000}. Such measurements promise to place exquisite constraints on the parameters governing these first galaxies that have hitherto been essentially unconstrained \citep{pober_et_al2014,greig_and_mesinger2015}. $21\,\textrm{cm}$ cosmology also has the potential to probe the pre-reionization epoch, providing insight into heating mechanisms in our early Universe, whether from X-ray sources \citep{christian_and_loeb2013,mesinger_et_al2014,ewall-wice_et_al2015}, structure formation shocks \citep{gnedin_and_shaver2004}, or exotic phenomena such as dark matter annihilation \citep{valdes_et_al2013,evoli_et_al2014}. From the standpoint of cosmology, a better understanding of reionization via $21\,\textrm{cm}$ cosmology removes a key nuisance from cosmic microwave background (CMB) datasets, allowing improved constraints on cosmological parameters such as the amplitude of primordial scalar fluctuations \citep{clesse_et_al2012,liu_et_al2015}. With the $21\,\textrm{cm}$ line potentially capable of accessing larger fractions of our observable Universe than any other cosmological probe, futuristic surveys will not only further refine our limits on fundamental parameters \citep{mcquinn_et_al2006,mao_et_al2008,visbal_et_al2009}, but may also provide precision tests of the inflationary paradigm \citep{cooray_et_al2008,barger_et_al2009}. For reviews of the capabilities of $21\,\textrm{cm}$ cosmology, see e.g. \citet{furlanetto_et_al2006,morales_and_wyithe2010,pritchard_and_loeb2012,aviBook}.

Recent years have seen increased experimental activity within $21\,\textrm{cm}$ cosmology. Current observational efforts targeting the reionization epoch can be roughly divided into two categories. Large radio interferometers such as the Murchison Widefield Array (MWA; \citealt{tingay_et_al2013,bowman_et_al2012}), the Donald C. Backer Precision Array for Probing the Epoch of Reionization (PAPER; \citealt{parsons_et_al2010}), the Low Frequency Array (LOFAR; \citealt{van_haarlem_et_al2013}), the Giant Metrewave Radio Telescope (GMRT; \citealt{paciga_et_al2013}), the Canadian Hydrogen Intensity Mapping Experiment (CHIME; \citealt{bandura_et_al2014}), and the proposed Square Kilometre Array (SKA; \citealt{mellema_et_al2013}) aim to measure the spatial fluctuations of the $21\,\textrm{cm}$ brightness temperature field, particularly the power spectrum of these fluctuations. In contrast, smaller scale experiments consisting of single dipole antennas or a small handful of elements seek to characterize the mean $21\,\textrm{cm}$ signal (the ``global signal") as a function of redshift \citep{shaver_et_al1999}. Examples of global signal experiments include the Experiment to Detect the Global Epoch of Reionization Signature (EDGES; \citealt{bowman2010}), the Large-Aperture Experiment to Detect the Dark Ages (LEDA; \citealt{greenhill2012}), the Shaped Antenna Measurement of the Background Radio Spectrum (SARA; \citealt{patra_et_al2013}), the Sonda Cosmol\'{o}gica de las Islas para la Detecci\'{o}n de Hidr\'{o}geno Neutro (SCI-HI; \citealt{voytek2014}), the Zero-spacing Interferometer Measurements of the Background Radio Spectrum (ZEBRA; \citealt{mahesh_et_al2014}), and the Broadband Instrument for Global Hydrogen Reionisation Signal (BIGHORNS; \citealt{Sokolowski_et_al2015}).

Multiple instruments have steadily increased in sensitivity while slowly overcoming foreground contaminants (such as Galactic synchrotron radiation), which are typically more than $10^4$ times brighter in temperature than theoretical expectations for the reionization $21\,\textrm{cm}$ signal \citep{dimatteo_et_al2002,santos_et_al2005,wang_et_al2006,deOliveiraCosta_et_al2008}. While a positive detection of the cosmological $21\,\textrm{cm}$ signal has yet to be made at the low frequencies corresponding to reionization redshifts (see \citealt{masui_et_al2013} and \citealt{switzer_et_al2013} for a detection at $z\sim0.8$ in cross-correlation with galaxy surveys), scientifically interesting limits are beginning to appear. \citet{paciga_et_al2013}, \citet{parsons_et_al2014}, \citet{dillon_et_al2014}, \citet{jacobs_et_al2015}, and \citet{dillon_et_al2015} have all set increasingly stringent upper limits on the $21\,\textrm{cm}$ power spectrum, and the current best limits \citep{ali_et_al2015} have been able to place constraints on IGM heating at $z \sim 8.4$ \citep{pober_et_al2015,greig_et_al2015a}. Global $21\,\textrm{cm}$ signal experiments have shown similar progress, with various published constraints on the strength of the signal \citep{rogers_and_bowman2008,bowman_et_al2008,voytek2014,vedantham_et_al2015}. The current best published results rule out the possibility of reionization happening more quickly than $\Delta z = 0.06$ \citep{bowman2010}.

Continual improvements in existing instruments (e.g., \citealt{mozdzen_et_al2015}) and the advent of new instruments make it likely for there to be a positive detection of the cosmological $21\,\textrm{cm}$ signal soon. Beyond a first detection, the proposed Square Kilometre Array (SKA; \citealt{mellema_et_al2013}) and the recently commenced Hydrogen Epoch of Reionization Array (HERA; \citealt{pober_et_al2014}) are forecasted to characterize the $21\,\textrm{cm}$ power spectrum with high signal-to-noise. New instruments such as the Dark Ages Radio Explorer (DARE; \citealt{burns2012}) are similarly being proposed to target the global signal.

With high significance results expected soon from both the power spectrum and global signal experiments, it is timely to re-examine the forecasted performance of upcoming experiments. In this paper, we build on previous forecasting exercises such as \citet{pober_et_al2014} and \citet{greig_and_mesinger2015} and implement a number of improvements in the forecasting methodology. Previous studies have typically kept cosmological parameters fixed while varying the astrophysical parameters that govern reionization. Here, we vary all parameters simultaneously, using HERA as a worked example to see how well these parameters can be constrained by $21\,\textrm{cm}$ power spectrum observations. HERA is a recently commenced next-generation instrument possessing $54,000\,\textrm{m}^2$ of collecting area, consisting of 331 drift-scanning dishes in a tightly packed hexagonal configuration to maximize sensitivity to the power spectrum. (For more details, see Section \ref{sec:methods}). The fiducial parameters used in our forecasts are chosen to match the latest CMB results from the \emph{Planck} satellite \citep{Planck2015overview,Planck2015parameters}. This is in contrast to previous studies, which typically centred their cosmological parameters on the best-fit values from the Wilkinson Microwave Anisotropy Probe (WMAP; \citealt{hinshaw_et_al2013}). Our updated forecasts thus capture the fact that \emph{Planck} predicts a lower redshift of reionization than WMAP did. We also show how power spectrum measurements can be paired with semi-analytic simulations of reionization to predict the ionization history, and forecast the expected uncertainties within this framework. A special emphasis is placed on how global signal and power spectrum measurements may be combined. We explore the power of joint fits and additionally propose a method for parameterizing the global signal based on power spectrum results.

The rest of this paper is organized as follows. In Section \ref{sec:ForecastUpdate} we provide our updated forecasts for astrophysical and cosmological parameter constraints, as well as our forecasts for ionization history constraints. Section \ref{sec:BruteJointFit} adds global signal measurements to the parameter fits. In Section \ref{sec:PspecHelpingGlobalSig} we show how parameter constraints from a power spectrum measurement can be used to construct a parameterization of the global signal that economically accounts for deviations from a fiducial global signal model with only a small handful of parameters. We summarize our conclusions in Section \ref{sec:Conclusions}.

\section{Updated constraints from power spectrum measurements}
\label{sec:ForecastUpdate}

In this section, we briefly update previous forecasts on the parameter estimation performance of upcoming instruments. Using HERA as a worked example, we will emphasize several points. First, we will show that despite the exquisite precision of base $\Lambda$CDM parameters from \emph{Planck}, errors on cosmological parameters have a non-negligible impact on parameter forecasts from $21\,\textrm{cm}$ power spectrum measurements. We will also find that while HERA is capable of producing precise parameter constraints over a wide range of reionization scenarios, those constraints are markedly more precise if reionization occurs at the lower end of the redshift range favoured by \emph{Planck}. Finally, we will generate forecasts for ionization history constraints based on a HERA measurement of the power spectrum. These forecasts will likely shift over the next few years, as early measurements allow for further refinement of underlying models of reionization.

\subsection{Forecasting methodology}
\label{sec:methods}
Fitting observational data (grouped here into a vector $\mathbf{d}$) with a model is tantamount to determining the likelihood function $\mathcal L(\mathbf{d}, \boldsymbol \theta)$, which can be viewed as a probability distribution function for the model parameters (grouped into a vector $ \boldsymbol \theta$). Forecasting the performance of an instrument is then roughly equivalent to determining the width of the likelihood function. For speed and simplicity, we perform this forecasting using the Fisher matrix formalism \citep{fisher1935}. The Fisher formalism approximates the likelihood function $\mathcal{L}$ as a multi-dimensional Gaussian in the parameters, with the Fisher matrix $\mathbf{F}$ quantifying the curvature of the negative logarithm of the likelihood, i.e.,
\begin{equation}
\label{eq:FisherDef}
\mathbf{F}_{\alpha \beta} = - \Bigg{\langle} \frac{\partial^2 \ln \mathcal{L}}{\partial \theta_\alpha \partial \theta_\beta} \Bigg{\rangle},
\end{equation}
where $\theta_\alpha$ is the $\alpha$th parameter that we wish to measure, and the angle brackets $\langle \dots \rangle$ denote an ensemble average over data. It is understood that the derivatives are to be evaluated about fiducial parameter values. The predicted error bar on $\theta_\alpha$ about its fiducial value is then given by $(\mathbf{F}^{-1})_{\alpha\alpha}^{1/2}$ if all parameters are fit jointly, and $(\mathbf{F}_{\alpha \alpha})^{-1/2}$ if all other parameters are held fixed \citep{kenney_and_keeping1951,kendall_and_stuart_1969,tegmark_et_al1997}. Because the Fisher matrix formalism assumes Gaussian likelihoods, it is an approximation to a full computation of the likelihood (e.g., using Markov Chain Monte Carlo methods). However, the Fisher formalism has the advantage of a much lower computational cost. This is crucial for the present work, for while computations of the full likelihood are feasible if only astrophysical parameters are varied \citep{greig_and_mesinger2015}, here we wish to simultaneously vary cosmological parameters. As long as the error bars are reasonably tight, the Fisher matrix approach should be an excellent approximation.

To compute the Fisher matrix for a measurement of the $21\,\textrm{cm}$ power spectrum, we follow the steps used in \citet{pober_et_al2014}. For a power spectrum measurement, Eq.~\eqref{eq:FisherDef} reduces to
\begin{equation}
\label{eq:FisherPspecDef}
\mathbf{F}^\textrm{PS}_{\alpha \beta}  = \sum_{k,z} \frac{1}{\varepsilon^2_\textrm{PS} (k,z)} \frac{\partial \Delta^2 (k,z)}{\partial \theta_\alpha} \frac{\partial \Delta^2 (k,z)}{\partial \theta_\beta},
\end{equation}
where the superscript serves to remind us that this pertains to a power spectrum measurement, $\varepsilon_\textrm{PS} (k,z)$ is the error (at spatial wavenumber $k$ and redshift $z$) on the measurement of $\Delta^2 (k,z) \equiv (k^3 / 2\pi^2) P_{21}(k,z)$, and the power spectrum $P_{21}(k,z)$ is defined as
\begin{equation}
\label{eq:PspecDef}
\langle  \widetilde{\Delta T}_{21} (\mathbf{k}, z)  \widetilde{ \Delta T}_{21}^* (\mathbf{k}^\prime, z) \rangle \equiv (2\pi)^3 \delta^D (\mathbf{k} - \mathbf{k}^\prime) P_{21}(k,z),
\end{equation}
with $\delta^D$ being a Dirac delta function and $\widetilde{\Delta T}_{21} (\mathbf{k},z)$ denoting the spatial Fourier transform\footnote{We adopt a Fourier convention where the forward transform is given by $\widetilde{f} (\mathbf{k}) \!=\! \int  f(\mathbf{r}) e^{-i \mathbf{k} \cdot \mathbf{r}} d^3\mathbf{r} $ and the inverse Fourier transform is given by $f(\mathbf{r}) \!=\! \int  \widetilde{f} (\mathbf{k}) e^{i \mathbf{k} \cdot \mathbf{r}}  \frac{d^3\mathbf{k}}{(2\pi)^3}$.} of $\Delta T_{21} (\mathbf{r},z)$, the spatial fluctuations of the $21\,\textrm{cm}$ brightness temperature contrast $T_\textrm{21}$ with respect to the CMB. On the right hand side of Eq.~\eqref{eq:PspecDef}, we have written $P_{21}(k,z)$ rather than $P_{21}(\mathbf{k},z)$, under the assumption that the cosmological signal is statistically isotropic. In practice, this assumption enters when estimates of the power spectrum are averaged in bins of constant $k = | \mathbf{k} |$. Note that in principle, the scatter in power within each such bin contains information, since the sample variance is proportional to the power spectrum signal itself. Our expression for the Fisher matrix does not include this information contribution, since it has been shown to be small \citep{ewall-wice_et_al2015}.

Calculating the derivatives of the power spectrum for Eq. \eqref{eq:FisherPspecDef} requires a model of reionization. Throughout this paper, we use the publicly available {\tt 21cmFAST} package \citep{mesinger_et_al2011} to simulate reionization, which employs the excursion set formalism of \citet{furlanetto_et_al2004} to compute ionization fields from simulated density fields. The package then uses the density, velocity, and ionization fields to compute the $21\,\textrm{cm}$ brightness temperature contrast against the CMB using \citep{wouthuysen1952,field1958,field1959,field1972,furlanetto_et_al2006,aviBook}
\begin{eqnarray}
\label{eq:deltaTdef}
T_{21}(\mathbf{\hat{n}}, \nu) \approx 27 x_\textrm{HI}  (1 + \delta_b) \left( 1 - \frac{T_\textrm{CMB}}{T_s}\right) \left( \frac{1}{1+H^{-1} \partial v_r / \partial r} \right) \nonumber \\
 \times  \left( \frac{1+z}{10} \frac{0.14}{\Omega_m h^2} \right)^\frac{1}{2} \left( \frac{\Omega_b h^2}{0.022}\right)\,\textrm{mK},
\end{eqnarray}
where $x_\textrm{HI}$ is the neutral fraction of the hydrogen atoms, $T_{21}$ is the temperature contrast against the CMB, $\delta_b$ is the baryon overdensity, $T_\textrm{CMB}$ is the CMB temperature, $T_s$ is the spin temperature, $\Omega_m$ is the normalized matter density, $\Omega_b$ is the normalized baryon density, $h$ is the Hubble parameter (in units of $100\,\textrm{km}/\textrm{s}/\textrm{Mpc}$), and $\partial v_r / \partial r$ is the derivative of the comoving radial peculiar velocity with respect to the comoving radial distance. During the reionization epoch, the IGM temperature is expected to be coupled to the spin temperature $T_s$, and is often assumed to have been pre-heated to a high temperature (from, say, early sources of X-rays or structure formation shocks). It is thus common to assume that $T_s \gg T_\textrm{CMB}$ \citep{hogan_and_rees1979,scott_and_rees1990,madau_et_al1997,zaldarriaga_et_al2004}, and with this approximation the spin temperature drops out of Eq. \eqref{eq:deltaTdef}. This significantly shortens the computational time of {\tt 21cmFAST}, since the evolution of $T_S$ is in general complicated to compute. With this paper focused on the reionization epoch, we make the $T_s \gg T_\textrm{CMB}$ assumption throughout. Once the $T_{21}$ field is obtained, the global signal $\overline{T}_{21}$ can be easily computed by averaging over all sky angles. Subtracting this mean signal from the simulated brightness temperature fields, the power spectrum can then be found using Eq. \eqref{eq:PspecDef}.

To compute the error $\varepsilon (k,z)$ of our power spectrum measurement, we make use of the {\tt 21cmSense} code\footnote{\href{https://github.com/jpober/21cmSense}{\tt https://github.com/jpober/21cmSense}} \citep{parsons_et_al2012a,pober_et_al2013,pober_et_al2014}. We assume that HERA consists of a tightly packed hexagonal array of $331$ dishes, each with a diameter of $14\,\textrm{m}$. The dishes are non-pointing, and thus all HERA observations will be drift-scan surveys. HERA is optimized for measurements between $100$ and $200\,\textrm{MHz}$, although there are plans to extend the lower end of this range. For the power spectrum forecasts in this paper, we assume $1080\,\textrm{hrs}$ of observations. To account for foreground contamination, we employ the ``moderate foregrounds" setting of {\tt 21cmSense}. This setting assumes that foregrounds are spectrally smooth, and thus can be mostly eliminated by filtering out low $k_\parallel$ spatial modes, where $k_\parallel$ is the spatial wavenumber along the line-of-sight (frequency) axis of observations. The threshold value of filtered $k_\parallel$ modes is baseline dependent, with more modes filtered out from data from long baselines. This is because longer baselines tend to be more chromatic, in the following sense. Suppose one were to examine the fringe pattern of a particular baseline as a function of frequency. As the fringe dilates/contracts with decreasing/increasing frequency, a given point in the sky is sampled by different phases of the fringe, with long baselines going through more cycles of the fringe pattern.  This cycling has the effect of causing sources that are intrinsically spectrally smooth (such as foregrounds) to appear chromatic in the data. Said differently, smooth foreground power leaks from low $k_\parallel$ modes to high $k_\parallel$ modes. Long baselines are more susceptible to this, and thus require a more conservative cut in Fourier space. For details, please see \cite{Datta2010,Parsons_et_al2012b,Morales2012,Vedantham2012,Trott2012,Thyagarajan2013,pober_et_al2014,Liu_et_al2014a,Liu_et_al2014b,Thyagarajan_et_al2015a,Thyagarajan_et_al2015b}.

In the parameter forecasts that follow, we will explore two fiducial scenarios. The first scenario is based on the best-fit cosmological parameters from \emph{Planck's} TT + lowP dataset. The astrophysical parameters needed to run the reionization simulation are then chosen so that when {\tt 21cmFAST} is run with the best-fit cosmological parameters, the resulting density and ionization fields predict an optical depth $\tau$ to the CMB that matches the best-fit $\tau$ from the dataset. We follow a similar procedure for \emph{Planck's} TT,TE,EE + lowP + lensing + ext dataset. With these two datasets, we roughly encompass the range of datasets in \citet{Planck2015parameters}: TT + lowP has some of the larger parameter errors and one of the highest redshifts of reionization, while TT,TE,EE + lowP + lensing + ext has some of the tightest errors and one of the lowest redshifts of reionization. When analyzing simulated data from the TT+lowP dataset, we assume that the data span $7.5 < z < 13$ and is divided into redshift bins of width $(\Delta z)_\textrm{bin} = 0.5$ before power spectra are generated separately for each bin; for the TT,TE,EE + lowP + lensing + ext dataset we assume a range of $6.1 < z < 13$ divided into bins of width $(\Delta z)_\textrm{bin} = 0.3$. In determining these specifications, the redshift ranges were mostly arbitrary, with the final results changing negligibly if we simply use the full redshift range of HERA for both datasets. Determining the bin widths is more subtle. Thicker bins mean that larger survey volumes go into the computation of each power spectrum, reducing both cosmic variance and instrumental noise errors. However, thicker bins also mean a more coarse-grained picture of how the power spectrum evolves with redshift, and previous works have shown that redshift evolution is a powerful way to break astrophysical parameter degeneracies \citep{pober_et_al2014,greig_and_mesinger2015}. There is thus an optimization problem to be solved, but we leave a definitive treatment of this to future work, as it would require a detailed treatment of light-cone effects \citep{barkana_and_loeb2006,datta_et_al2012,datta_et_al2014,zawada_et_al2014,laplante_et_al2014}, which are neglected in this paper. Our choices of $(\Delta z)_\textrm{bin} = 0.3$ and $(\Delta z)_\textrm{bin} = 0.5$ were motivated by simple numerical tests that indicated reasonably good performance, and are not to be taken as the results of a rigorous optimization.

\subsection{Astrophysical and cosmological parameter constraints}
\label{eq:PlainParameterConstraints}

\begin{table*}
\caption{\label{tab:Params} Parameter estimation forecasts for astrophysical and cosmological parameters. All quoted errors represent marginalized $68\%$ intervals. The columns labeled ``Errors" list the error bars obtained from the \emph{Planck} satellite \citep{Planck2015overview,Planck2015parameters}. The ``$+P_{21} (k)$" columns show the effect of adding $21\,\textrm{cm}$ power spectrum data from HERA, while the ``$+T_{21}^8(\nu)$" columns additionally include a global signal experiment requiring an eighth order polynomial foreground fit in $\ln \nu$. Adding $21\,\textrm{cm}$ power spectrum information to existing constraints results in reduced errors on many cosmological parameters (in addition to producing the first measurements of the astrophysical parameters governing reionization). The further addition of global signal information, however, does little to improve parameter errors. One also sees that the predicted errors are smaller when assuming fiducial values tied to the \emph{Planck} TT,TE,EE+lowP+lensing+ext dataset than when they take their \emph{Planck} TT+lowP values. The ``$+P_\textrm{21}(k)$" values in this table are reproduced in \citep{liu_et_al2015}, which builds on this paper to explore how a precise understanding of reionization from $21\,\textrm{cm}$ cosmology can improve CMB constraints.
}
\begin{tabular}{lllllcllll}
\hline\hline
 & \multicolumn{4}{c}{\textbf{\emph{Planck} TT + lowP}} && \multicolumn{4}{l}{\textbf{ \emph{Planck} TT,TE,EE+lowP+lensing+ext}  } \\
 & Best fit & Errors &  $+P_{21} (k)$ &$+T_{21}^8(\nu)$&& Best fit & Errors &  $+P_{21} (k)$ &  $+T_{21}^8(\nu)$\\
\hline
\multicolumn{7}{l}{Cosmological parameters} \\
$\Omega_b h^2$ \dotfill & $0.02222 $&$ \pm 0.00023$ & $\pm 0.00021$ &  $\pm 0.00021$ && $0.02230 $&$\pm 0.00014$ & $\pm 0.00013$ &  $\pm 0.00013$ \\
$\Omega_c h^2$ \dotfill & $0.1197$&$ \pm 0.0022$  & $\pm 0.0021$ &  $\pm 0.0021$ && $0.1188$&$ \pm 0.0010$ & $\pm 0.00096$ &  $\pm 0.00095$ \\
$100 \theta_\textrm{MC}$\dotfill  & $1.04085 $&$\pm 0.00046$ & $\pm 0.00046$ &  $\pm 0.00045$ & &$1.04093 $&$\pm 0.00030$ & $\pm 0.00029$ & $\pm 0.00029$ \\
$\ln ( 10^{10} A_s) \dots$ \dotfill & $3.089 $&$\pm 0.036$ & $\pm 0.023$ &  $\pm 0.022$ & &$3.064$&$ \pm 0.023$ & $\pm 0.016$ & $\pm 0.016$  \\
$ n_s $\dotfill  & $ 0.9655$&$ \pm 0.0062$ & $\pm 0.0057$ &  $\pm 0.0056$ && $0.9667$&$ \pm 0.0040$ & $\pm 0.0037$ & $\pm 0.0037$ \\
$ \tau $ \dotfill & $0.078 $&$\pm 0.019$ & $\pm 0.013$ & $\pm 0.013$  && $0.066 $&$\pm 0.012$ & $\pm 0.0089$ &  $\pm 0.0089$ \\
\hline
\multicolumn{7}{l}{Astrophysical parameters} \\
$T_\textrm{vir}\left[ \textrm{K} \right]$ \dotfill & $40000 $& --- & $\pm 7500$ & $\pm 7400$  && $60000 $& --- & $\pm 6700$ &  $\pm 6500$ \\
 $R_\textrm{mfp}\left[ \textrm{Mpc} \right]$\dotfill & $35.0 $& --- & $\pm 1.2$ & $\pm 1.2$  && $35.0 $& --- & $\pm 0.82$ &  $\pm 0.82$ \\
$\zeta$ \dotfill & $40.0 $& --- & $\pm 4.6$ & $\pm 4.5$  && $30.0 $& --- & $\pm 2.0$ &  $\pm 1.9$ \\
\hline\hline
\end{tabular}
\end{table*}

For our parameter estimation forecasts, we pick the parameter set $\boldsymbol \theta=[ \Omega_b h^2, \Omega_c h^2, \theta_\textrm{MC}, \ln( 10^{10} A_s), n_s, \tau, T_\textrm{vir}, R_\textrm{mfp}, \zeta]$, where $\Omega_c$ is the normalized cold dark matter density, $\theta_\textrm{MC}$ is the approximate angular size of the sound horizon at recombination (as defined by the {\tt CosmoMC} package; \citealt{lewis_and_bridle2002}), $A_s$ is the amplitude of primordial scalar fluctuations, $n_s$ is the scalar spectral index, and the rest of the cosmological parameters are as they were defined above.

We supplement the cosmological parameters with three astrophysical parameters. The first is $T_\textrm{vir}$. This is the minimum virial temperature for which a halo is expected to harbor substantial star formation. Typically, $T_\textrm{vir}$ is assumed to be at least $\sim 10^4\,\textrm{K}$, corresponding to the threshold at which efficient atomic hydrogen line cooling occurs. Note that our choice of a constant virial temperature means that the virial mass $M_\textrm{vir}$ is redshift-dependent, since the two quantities are related via (see, e.g., \citealt{barkana_and_loeb2001})
\begin{equation}
T_\textrm{vir} = 1.98 \times 10^4\,\textrm{K} \left( \frac{\mu}{0.6}\right) \left( \frac{M_\textrm{vir}}{10^8 h^{-1} M_\odot}\right)^\frac{2}{3} \left[ \frac{\Omega_m}{\Omega_m^z} \frac{\Delta_c}{18 \pi^2} \right]^\frac{1}{3} \left( \frac{1+z}{10} \right),
\end{equation}
where $\mu$ is the mean molecular weight, $\Delta_c \equiv 18 \pi^2 + 82 d - 39 d^2$, $d\equiv \Omega_m^z -1$, and
\begin{equation}
\Omega_m^z \equiv \frac{\Omega_m (1+z)^3}{\Omega_m (1+z)^3+\Omega_k (1+z)^2+\Omega_\Lambda},
\end{equation}
with $\Omega_k$, $\Omega_m$, and $\Omega_\Lambda$ being the normalized curvature, matter, and dark energy densities, respectively. Our second astrophysical parameter is $R_\textrm{mfp}$. This is defined as the mean free path of ionizing photons in ionized regions. Photons in ionized regions cannot travel arbitrarily large distances because the ionized regions are not \emph{completely} ionized---dense self-shielded pockets of neutral hydrogen reside inside the galaxies that created the ionized regions in the first place. Ionizing photons thus have a limited range even inside ionized regions, and in practice this sets a upper limit on the bubble size in {\tt 21cmFAST} simulations. Our final astrophysical parameter is $\zeta$, the ionizing efficiency. This can be written as
\begin{equation}
\zeta \equiv f_* f_\textrm{esc} N_{\gamma / b} (1+n_\textrm{rec})^{-1},
\end{equation}
where $f_*$ is the star formation efficiency, $f_\textrm{esc}$ is the escape fraction of ionizing photons from galaxies, $N_{\gamma / b}$ is the number of ionizing photons produced per baryon in stars, and $n_\textrm{rec}$ is the average number of recombinations. With this parameterization, an object of mass $m_\textrm{obj}$ is able to ionize a region enclosing mass $\zeta m_\textrm{obj}$. We stress that the three astrophysical parameters presented here are not ``fundamental" in any way, and are simply designed to capture variations in current semi-analytic models of reionization. It is likely that as the first detections of the $21\,\textrm{cm}$ reionization signal become available that both the models and their parameterizations will evolve. We thus do not recommend taking the uncertainties on $T_\textrm{vir}$, $R_\textrm{mfp}$, and $\zeta$ (presented below) as the final word on the precision to which reionization may be understood, though they may be useful as a preliminary guide.

\begin{figure*}
\centering
\includegraphics[width=1.0\textwidth]{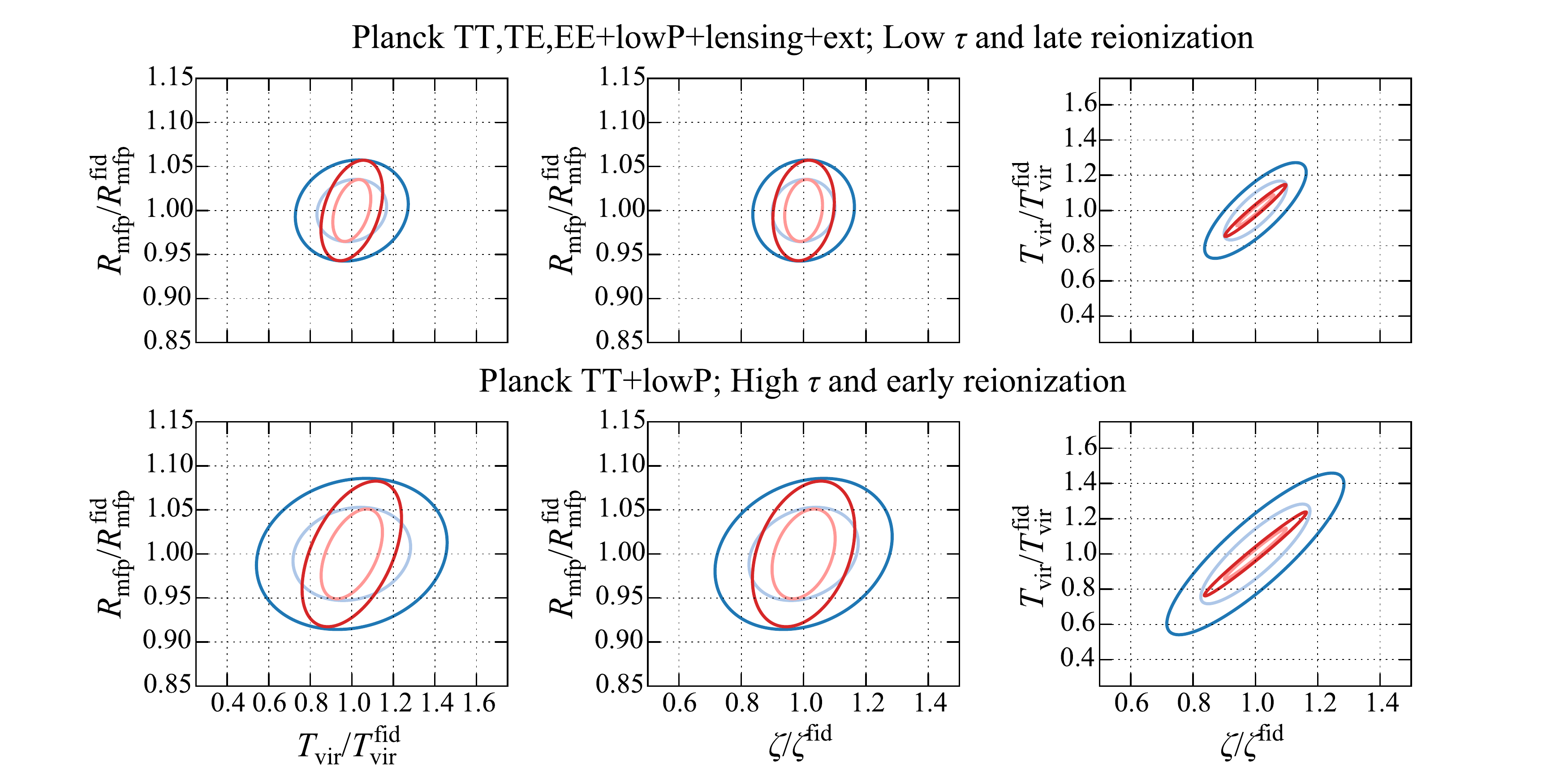}
\caption{Forecasted astrophysical parameter constraints from HERA. Light contours signify $68\%$ confidence regions, while dark contours denote $95\%$ confidence regions. Axes are scaled according to fiducial values chosen to match \emph{Planck's} TT,TE,EE+lowP+lensing+ext data (top row) and \emph{Planck's} TT+lowP data (bottom row). Red contours assume that cosmological parameters are known, whereas blue contours marginalize over cosmological parameter uncertainties. Though the CMB (in conjunction with other cosmological probes) has delivered exquisite constraints on cosmological parameters, we see that the residual uncertainties still have a non-negligible effect on astrophysical parameters derived from $21\,\textrm{cm}$ power spectrum measurements. When only astrophysical parameters are varied (red contours), the constraints are in rough agreement with those found in \citet{pober_et_al2014}, although direct comparisons should be interpreted with caution given the different priors on fiducial parameters.}
\label{fig:cosmoVsAstro}
\end{figure*}

Table \ref{tab:Params} shows the parameter errors from \emph{Planck} (columns labeled ``Errors"; \citealt{Planck2015parameters}) and the improved errors resulting from adding $21\,\textrm{cm}$ power spectrum measurements from HERA (columns labeled ``$+P_{21} (k)$"). All quoted errors correspond to $68\%$ confidence regions after having marginalized over all other parameters.\footnote{For reasons of numerical stability, we treat the $\theta_\textrm{MC}$ parameter a little differently from the other cosmological parameters. We assume that the $21\,\textrm{cm}$ power spectrum does not directly constrain $\theta_\textrm{MC}$. This means that any reductions in the errors on $\theta_\textrm{MC}$ come from breaking degeneracies from other parameters, and not from direct measurements. In practice, this has a negligible effect on the results, since $\theta_\textrm{MC}$ is determined to such high precision from CMB results alone. We do the same for the global signal constraints on $R_\textrm{mfp}$ computed in Section \ref{sec:BruteJointFit}.} Note that we assume all parameters (whether cosmological or astrophysical) are varied simultaneously; this is in contrast to \citet{pober_et_al2014}, which only dealt with the astrophysical parameters and assumed that cosmological parameters were known. We find that such a simultaneous variation of parameters is important even if one is primarily interested in only the astrophysical parameters. This is illustrated in Fig. \ref{fig:cosmoVsAstro}, where pairwise parameter constraints on the three reionization astrophysics parameters, $T_\textrm{vir}$, $R_\textrm{mfp}$, and $\zeta$ are shown. Dark contours delineate $95\%$ confidence regions, while light contours delineate $68\%$ confidence regions. Red contours show the constraints that can be obtained if cosmological parameters are known perfectly, whereas blue contours show the degraded constraints that result from marginalizing over \emph{Planck}-level spreads in the cosmological parameters. The $95\%$ confidence regions of the former are often roughly the same size as the $68\%$ confidence regions of the latter, making it clear that cosmological parameter uncertainties are non-negligible.

Examining the values in Table \ref{tab:Params}, one sees that $21\,\textrm{cm}$ power spectrum measurements are more effective at constraining reionization astrophysics if the best-fit parameters have a lower $\tau$ value, implying a lower redshift of reionization (as is the case for TT,TE,EE + lowP + lensing + ext compared to TT + lowP). If reionization happens at lower redshifts, characteristic features in the evolution of the $21\,\textrm{cm}$ power spectrum (such as the peak in power at $50\%$ ionization fraction \citealt{lidz_et_al2008,bittner_and_loeb2011}) occur at higher frequencies, where both foregrounds and instrumental noise are lower. This speaks to a natural complementarity between the CMB and $21\,\textrm{cm}$ cosmology when it comes to constraining reionization. For example, if reionization occurs at high redshifts, $21\,\textrm{cm}$ measurements become harder. But $\tau$ is then larger, and the ``reionization bump" feature of CMB polarization power spectra \citep{zaldarriaga1997} becomes easier to measure, reducing the errors on CMB-derived reionization constraints.

\begin{figure*}
\centering
\includegraphics[width=1.0\textwidth,trim=0cm 3cm 2cm 3cm,clip]{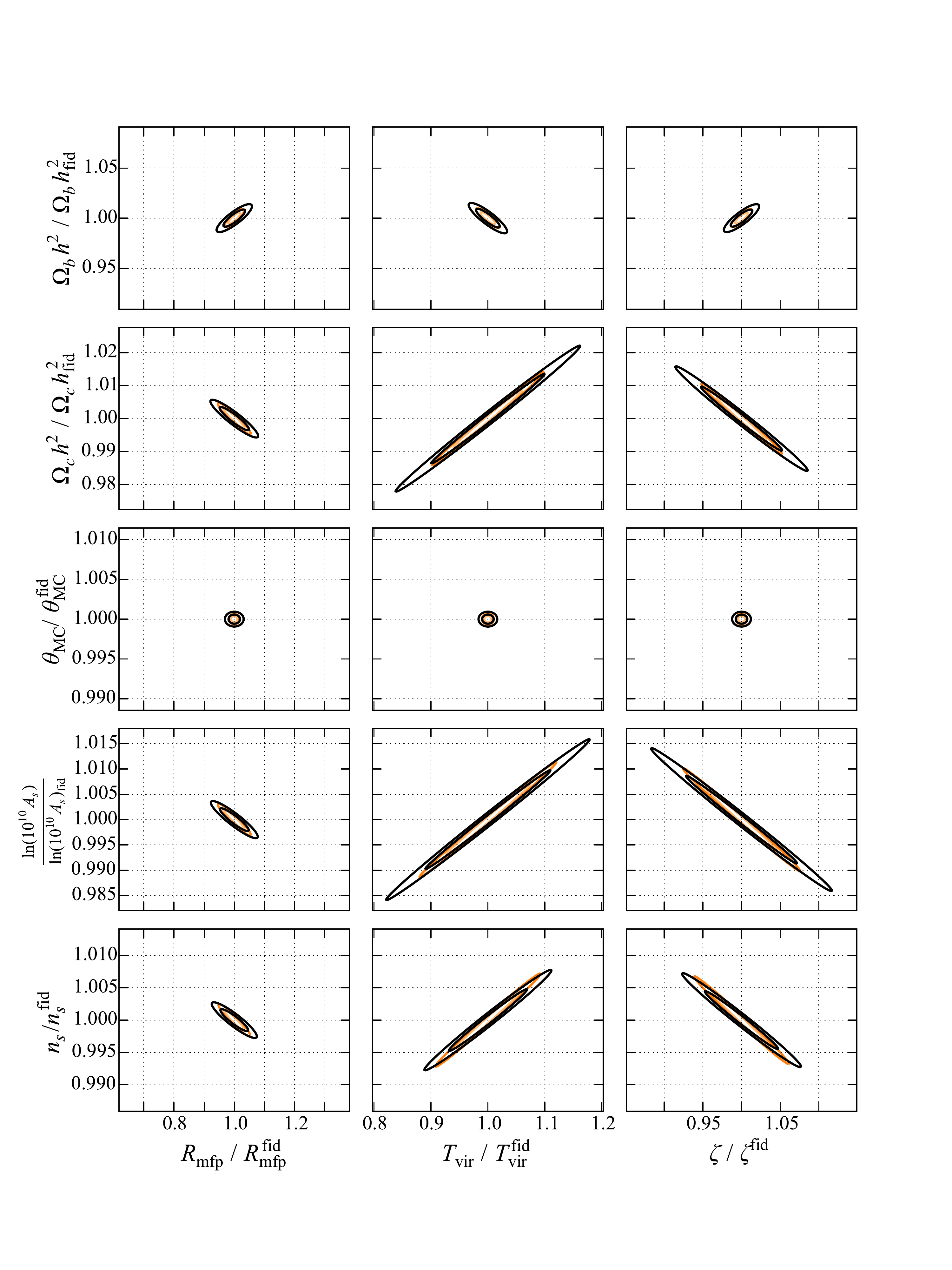}
\caption{Pairwise parameter contours between astrophysical and cosmological parameters for HERA, \emph{assuming all other parameters are held constant}. Light contours signify $68\%$ confidence regions, while dark contours denote $95\%$ confidence regions. Black contours show the results for fiducial parameters chosen to match \emph{Planck's} TT+lowP dataset, while orange contours show the results for \emph{Planck's} TT,TE,EE+lowP+lensing+ext.}
\label{fig:cosmoAstroDegensNoMarginalization}
\end{figure*}

\begin{figure*}
\centering
\includegraphics[width=1.0\textwidth,trim=0cm 3cm 2cm 3cm,clip]{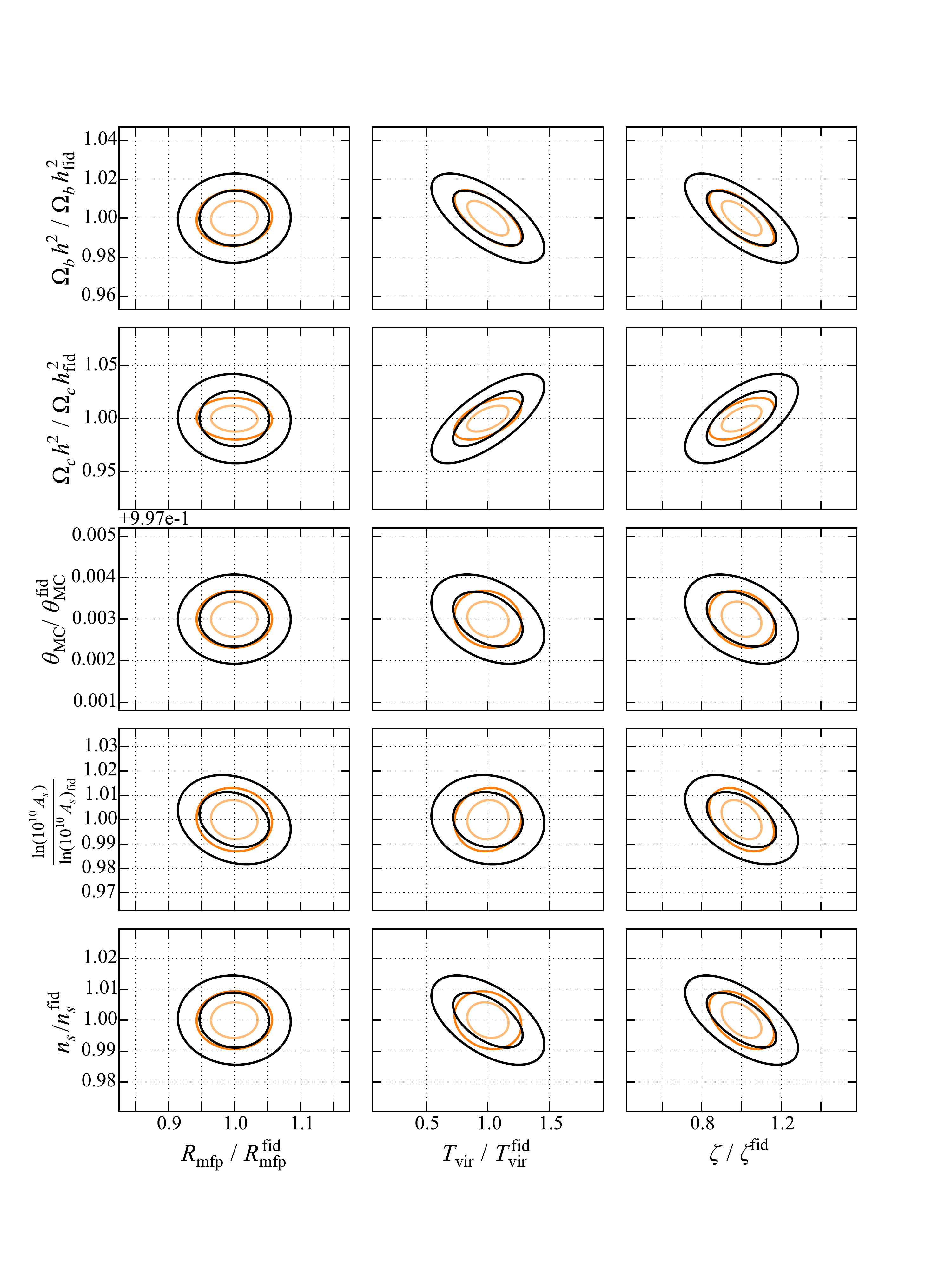}
\caption{Same as Fig. \ref{fig:cosmoAstroDegensNoMarginalization}, but with each pairwise plot showing parameter constraints assuming that all other parameters are marginalized over (rather than assumed to be fixed).}
\label{fig:cosmoAstroDegensWithMarginalization}
\end{figure*}

For the cosmological parameter constraints, the benefits from adding $21\,\textrm{cm}$ power spectrum measurement clearly vary from parameter to parameter. For example, errors in $\Omega_b h^2$, $\Omega_c h^2$, and $\theta_\textrm{MC}$ are essentially unchanged, while $A_s$, $n_s$, and $\tau$ do show some improvement. The origins of these improvements can be understood by examining Figs. \ref{fig:cosmoAstroDegensNoMarginalization} and \ref{fig:cosmoAstroDegensWithMarginalization}. Both figures show pairwise parameter constraints between astrophysical and cosmological parameters, but in Fig. \ref{fig:cosmoAstroDegensWithMarginalization} the parameters that are omitted from each pairwise plot are marginalized over, whereas in Fig. \ref{fig:cosmoAstroDegensNoMarginalization} the omitted parameters are held fixed. Though it is somewhat artificial to hold parameters fixed (when they are in fact quite uncertain), this has the advantage of providing a cleaner sense for the underlying physics. One sees particularly tight degeneracies between the astrophysical parameters and $A_s$, $n_s$, and $\Omega_c h^2$.

Physically, a higher $A_s$ means that density peaks will more quickly reach the threshold densities necessary for the collapse of overdensities into objects, pushing galaxy formation to higher redshifts. Increasing $\Omega_c h^2$ has a similar effect. From the perspective of a $21\,\textrm{cm}$ power spectrum measurement, one can compensate for this earlier structure formation by correspondingly reducing the ionizing efficiency. This means that early galaxies produce fewer ionizing photons, leaving the timing of reionization largely unchanged. There is thus a negative correlation between $A_s$ and $\zeta$, as well as between $\Omega_c h^2$ and $\zeta$. The correlations between $T_\textrm{vir}$ and cosmological parameters are similar, except with the opposite sign because the timing of reionization can be held constant even with earlier structure formation from higher $A_s$ or $\Omega_c h^2$ if reionization is driven by higher mass---and therefore rarer---halos with higher $T_\textrm{vir}$.

In a similar way to $A_s$, a change in $n_s$ can change the amount of structure formation on scales relevant to ionizing galaxies. Since such scales are at much higher $k$ than the pivot scale of a power-law primordial matter power spectrum, increasing $n_s$ is qualitatively similar to an increase in $A_s$. One therefore sees roughly the same behaviour in the last two rows of Fig. \ref{fig:cosmoAstroDegensNoMarginalization}, corresponding to changes in  $A_s$ and $n_s$.

Finally, consider $\Omega_b h^2$ and $\theta_\textrm{MC}$. In our {\tt 21cmFAST} runs, changing $\Omega_b h^2$ mainly affects our results through the dependence of the $21\,\textrm{cm}$ brightness temperature on $\Omega_b h^2$ that is seen in Eq. \eqref{eq:deltaTdef}. This may change with a more sophisticated prescription for the bias of the neutral hydrogen distribution. As for $\theta_\textrm{MC}$, we find that \emph{Planck} constrains this parameter so well that variations within its allowed range have very little impact on the $21\,\textrm{cm}$ results.

While Fig. \ref{fig:cosmoAstroDegensNoMarginalization} was helpful for guiding physical intuition, Fig. \ref{fig:cosmoAstroDegensWithMarginalization} provides a more realistic sense for parameter errors, since one would need to marginalize over unknown parameters in practice. With Fig. \ref{fig:cosmoAstroDegensWithMarginalization}, the trends are less obvious. Nonetheless, one continues to see covariances between the parameters that govern astrophysics and those that govern cosmology. Reasoning backwards, it is therefore unsurprising that we saw in Table \ref{tab:Params} that measurements of the $21\,\textrm{cm}$ power spectrum can provide additional constraints on cosmological parameters, thus reducing their errors.

In closing, we emphasize that the improvement in errors on $\tau$ arises for a different reason than the improvements in the other cosmological parameters. In simulations of reionization, $\tau$ is typically not an independent parameter that can be varied; rather, it is a \emph{prediction} that can be extracted from the simulated density and ionization fields, given the other model parameters. Imposing self-consistency between this prediction and the best-fit $\tau$ value derived from CMB data can then provide an additional opportunity to improve constraints on both $\tau$ and the other parameters. In this paper, we do not impose this self-consistency condition, leaving a discussion of its impact to \citet{liu_et_al2015}. With our current treatment, $\tau$ errors are reduced with the addition of $21\,\textrm{cm}$ only because $A_s$ and $\tau$ obey a tight degeneracy when constrained from CMB data alone (with $A_s e^{-2\tau}$ measured with much higher precision than either $A_s$ or $\tau$ alone). Reducing $A_s$ errors thus reduces $\tau$ errors.

\subsection{Ionization history constraints}
\label{sec:ionHist}

\begin{figure}
\centering
\includegraphics[width=0.53\textwidth]{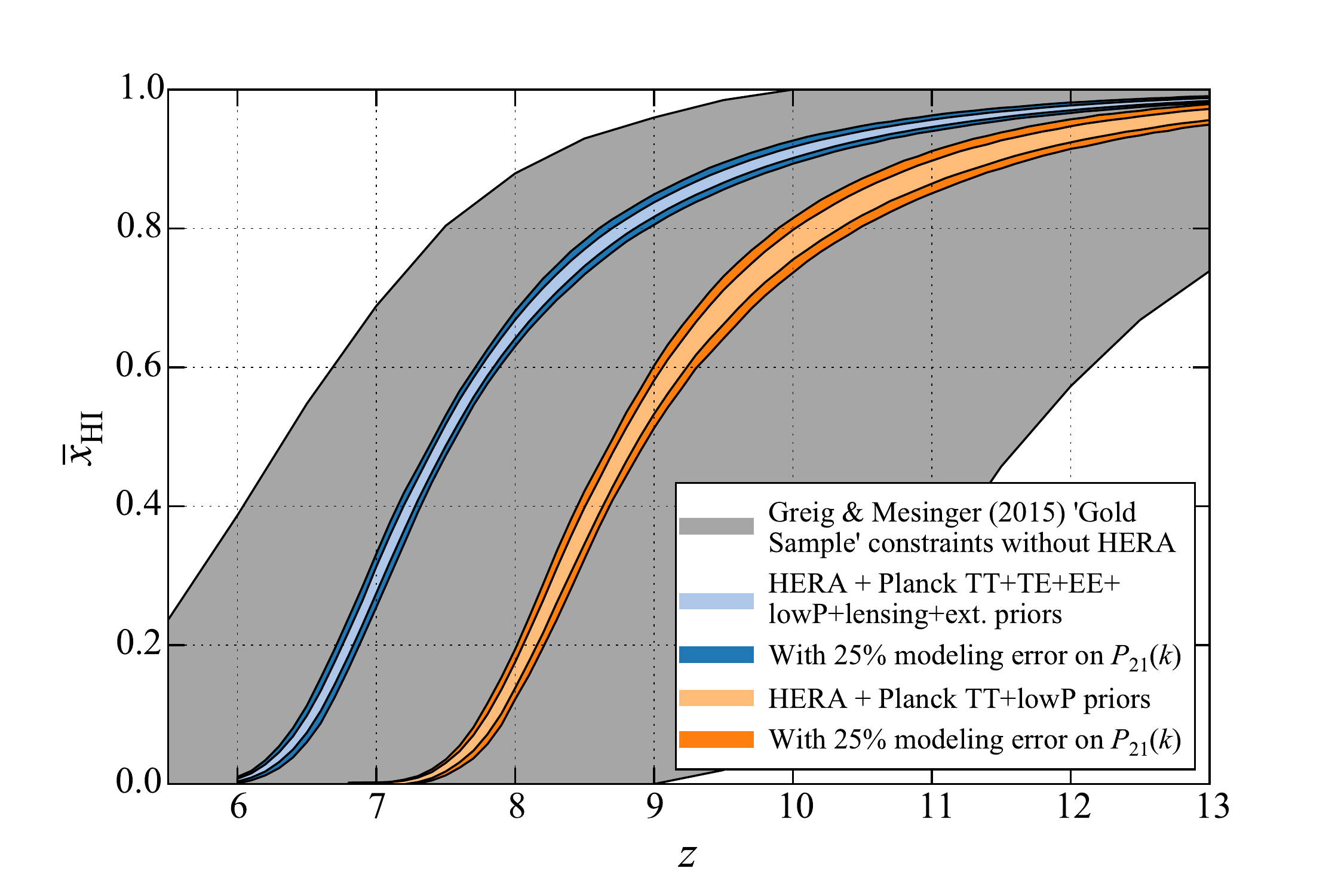}
\caption{Forecasts for constraints on the neutral fraction history $\overline{x}_\textrm{HI} (z)$ from HERA power spectrum measurements. A neutral fraction (or equivalently, ionization) history can be obtained from power spectrum measurements by first fitting the power spectra to predictions from semi-analytic reionization codes like {\tt 21cmFAST}, in the process obtaining best-fit values for the astrophysical and cosmological parameters listed in Table \ref{tab:Params}. These parameters can then be fed back into the code to predict the ionization history. The light blue band shows the expected $95\%$ confidence region for the neutral fraction history predicted from such a procedure, tuned to fit the fiducial parameters from \emph{Planck's} TT,TE,EE+lowP+lensing+ext dataset. The dark blue band shows the corresponding region if the semi-analytic codes are assumed to have a $25\%$ modeling error in their prediction of the $21\,\textrm{cm}$ power spectrum. Analogous constraints assuming fiducial values from \emph{Planck's} TT+lowP dataset are shown in orange. These forecasts compare favourably to the combined ``gold sample" constraints (grey band) from Greig \& Mesinger (in prep.).}
\label{fig:ionHist}
\end{figure}

The parameter constraints derived in the previous section can be translated into errors on the angle-averaged ionization history by drawing random realizations of parameters consistent with our predicted parameter covariances, and then running a simulation of the ionization field for each realized parameter set. The spread in the resulting ionization histories provides an estimate of the error in the ionization history from a power spectrum-based constraint.

In practice, even fast semi-analytic codes such as {\tt 21cmFAST} are insufficiently quick to allow this procedure to be followed if cosmological parameters are varied along with the astrophysical parameters. Assuming that errors about fiducial parameters are small, we instead adopt a linearized approach. Denoting by $\Delta p$ the shift in the parameter $p$ from its fiducial value, the angle-averaged neutral fraction $\overline{x}_\textrm{HI}$ is given by
\begin{equation}
\overline{x}_\textrm{HI} (z) = \overline{x}_\textrm{HI}^\textrm{fid}(z) + \sum_\gamma h_\gamma (z) \Delta p_\gamma ,
\end{equation}
where $x_\textrm{HI}^\textrm{fid}$ is the fiducial ionization history corresponding to the fiducial parameters (i.e. with $\Delta p_\gamma  = 0$), and $h_\gamma (z)$ is the redshift-dependent response of the ionization history to a shift in the $\gamma $th parameter (which we compute using training runs of {\tt 21cmFAST}).

Fig. \ref{fig:ionHist} shows the resulting ionization history constraints from HERA. In light orange is the $95\%$ confidence region assuming \emph{Planck} TT + lowP fiducial parameters, while the $95\%$ confidence region assuming \emph{Planck} TT,TE,EE + lowP + lensing + ext fiducial parameters is shown in light blue. In both cases, we simultaneously account for uncertainties in astrophysical and cosmological parameters, i.e., the confidence regions reflect uncertainties in all the parameters listed in Table \ref{fig:cosmoVsAstro} except $\tau$, which is not needed in the reionization simulations. Again, we see that the projected constraints are tighter if reionization happens at lower redshifts, but in either scenario HERA is able to make a precise measurement. Of course, this measurement is necessarily model-dependent, since the $21\,\textrm{cm}$ power spectrum measurement does not directly measure the ionization history. To account for the model dependence, the dark blue and dark orange band show the results from incorporating a modeling error in the power spectrum constraints.  We follow the treatment in \citet{greig_and_mesinger2015} and crudely add an uncorrelated modeling error to each $k$ bin at a level of $25\%$ of the fiducial power spectrum, in quadrature with the errors computed by {\tt 21cmSense}. This essentially incorporates a bin-by-bin amplitude error in our modeling of the power spectrum. The increased errors then result in greater astrophysical and cosmological parameter uncertainties, which translate into greater uncertainties in the ionization history. One sees that the final constraints remain tight. In addition to the amplitude errors, there also exists a generous allowance for possible modeling errors in the \emph{timing} of reionization, with our ionization history constraints remaining relevant even if broadened by, say, $\Delta z \sim 0.5$ or $1$. This can be seen by comparing the HERA constraints to the grey band in Fig. \ref{fig:ionHist}, which is reproduced from the combined ``gold sample" constraints of Greig \& Mesinger (in prep.). To obtain these constraints, Greig \& Mesinger used {\tt 21cmFAST} simulations within a Bayesian framework to combine constraints from the \emph{Planck} optical depth with those from dark pixels in the Ly$\alpha$ and Ly$\beta$ forests \citep{mcgreer_et_al2015}. The $21\,\textrm{cm}$ line is seen to have much to contribute.

\section{Combining power spectrum and global signal measurements via joint fits}
\label{sec:BruteJointFit}

Having provided updated power spectrum parameter constraints, we now consider how global signal and power spectrum measurements can complement one another. In this section, we examine what is arguably the simplest way to combine two complementary datasets---a joint fit for the same model parameters with both datasets.

Global signal measurements measure the $21\,\textrm{cm}$ brightness temperature averaged over the entire sky \citep{madau_et_al1997,shaver_et_al1999,gnedin_and_shaver2004,sethi2005,bowman_et_al2008,bowman2010,pritchard_and_loeb2010}. In other words, global signal experiments do not resolve the spatial variations of the brightness temperature field. The goal is instead to average over these variations to obtain an average brightness temperature $\overline{T}_\textrm{21}$ as a function of redshift (or equivalently, frequency). Typically, the global signal is accessed by a single antenna with a spatial response that is broad enough to essentially perform the requisite average over the sky (see \citealt{mahesh_et_al2014,vedantham_et_al2015,presley_et_al2015,singh_et_al2015,venumadhav_et_al2015} for discussions of some alternatives). As our worked example, we assume that this antenna is pointed towards the Northern Galactic Pole (NGP), with a beam pattern $A(\theta, \varphi)$ given by
\begin{equation}
A(\theta, \varphi) = \exp \left( -\frac{1}{2} \frac{\theta^2}{\theta_b^2} \right) \cos \theta,
\end{equation}
where $\theta$ is the polar angle from the NGP, $\varphi$ is the azimuthal angle, and we take $\theta_b$ to be $0.3\,\textrm{rad}$ at $120\,\textrm{MHz}$. We assume a frequency range of $120\,\textrm{MHz}$ to $200\,\textrm{MHz}$. To account for foreground contamination, a common approach is to fit the measured spectra with low order polynomials. The expectation is that foregrounds are typically spectrally smooth, and can therefore be accounted for by low-order fits \citep{shaver_et_al1999,pritchard_and_loeb2010,morandi_and_barkana2012,vedantham_et_al2014}. In this section, we use Legendre polynomials as an orthogonal basis for our fits. Explicitly, the foreground contribution $\overline{T}_\textrm{fg} (\nu)$ to the global signal is modeled as 
\begin{equation}
\label{eq:GlobalSigFG}
\overline{T}_\textrm{fg} (\nu) = \exp \left[ \sum_{i=0}^{N_p} a_i P_i \left(\frac{ \ln \nu - \ln \nu_0}{\Delta_{\ln \nu}} \right) \right],
\end{equation}
where $P_i$ is the $i$th Legendre polynomial, $\{ a_i \}$ are the foreground model parameters to be fit for in the data, and $N_p$ is the order of foreground polynomial. The quantities $\nu_0$ and $\Delta_{\ln \nu}$ are defined for convenience so that if $\nu_\textrm{min}$ and $\nu_\textrm{max}$ are the minimum and maximum observation frequency, respectively, we have $\ln \nu_0 \equiv (\ln \nu_\textrm{min} + \ln \nu_\textrm{max}) / 2$ and $\Delta_{\ln \nu} \equiv (\ln \nu_\textrm{max} - \ln \nu_\textrm{min}) / 2$. Following \citet{bernardi_et_al2015}, we assume that eighth order polynomials are sufficient to fit out the foregrounds, although in Section \ref{sec:PspecHelpingGlobalSig} we varying $N_p$ to examine the impact that foregrounds may have on a detection of the global signal.

To forecast the performance of a global signal experiment, we again employ the Fisher matrix formalism. The Fisher matrix is given by \citep{pritchard_and_loeb2010,bernardi_et_al2015,presley_et_al2015}
\begin{equation}
\mathbf{F}^\textrm{GS,FG}_{\alpha \beta} = \sum_i \frac{1}{\varepsilon^2_\textrm{GS}(\nu_i)} \frac{\partial \overline{T}_\textrm{tot}(\nu_i)}{\partial \phi_\alpha}\frac{\partial \overline{T}_\textrm{tot}(\nu_i)}{\partial \phi_\beta},
\end{equation}
where $ \overline{T}_\textrm{tot} \equiv \overline{T}_\textrm{fg} + \overline{T}_{21}$ is the total (foregrounds plus cosmological) signal measured by an instrument, and $\phi_\alpha$ is the $\alpha$th parameter in the set of model parameters $\boldsymbol \phi$, which consists of $\boldsymbol \theta$ plus the polynomial coefficients used to fit for foregrounds. The instrumental noise error in the measurement of $\overline{T}_\textrm{tot}$ is denoted by $\varepsilon_\textrm{GS}$. To combine the Fisher information from the global signal with that from power spectrum, we must first marginalize over the nuisance foreground coefficients. Operationally, this means inverting $\mathbf{F}^\textrm{GS,FG}$ to obtain a covariance matrix for the parameter set $\boldsymbol \phi$, deleting the rows and columns corresponding to the foreground coefficients, and inverting once more to obtain a smaller Fisher matrix $\mathbf{F}^\textrm{GS}$ for the parameter set $\boldsymbol \theta$. The final Fisher matrix for a joint fit between results from a power spectrum and a global signal measurement is then obtained by summing $\mathbf{F}^\textrm{PS}$ and $\mathbf{F}^\textrm{GS}$.

The parameter forecasts for a combined fit are shown in the ``$+T_{21}^8(\nu)$" column of Table \ref{tab:Params}. One sees that a global signal experiment targeting the reionization epoch (i.e., covering frequency ranges above $\sim 100\,\textrm{MHz}$) adds little to parameter constraints (whether astrophysical or cosmological) already provided by the power spectrum measurements. This is consistent with the results found in \citet{mirocha_et_al2015}, which highlighted the importance of broadband measurements that reach the high redshifts of the pre-reionization era. During reionization, the global signal is essentially smooth and featureless, and takes the simple form of a slow decay to zero brightness temperature as the neutral hydrogen fraction approaches zero (see, for instance, the solid black curves in Fig. \ref{fig:devEigenmodes}). With so few features to constrain, observations of the global signal during reionization are highly degenerate in their determination of model parameters, particularly when nuisance parameters from the spectrally smooth foregrounds are included. Instead of seeking constraints on physical model parameters, global signal measurements from reionization are far better suited for capturing broad phenomenological parameters such as the rough redshift and duration of reionization.

\section{Using a power spectrum measurement to aid a global signal measurement}
\label{sec:PspecHelpingGlobalSig}

In the previous section, we saw that joint fits of data from power spectrum and global signal experiments are unlikely to reduce errors in astrophysical and cosmological parameter constraints, at least if measurements are limited to the reionization epoch. Furthermore, joint fits implicitly assume a correct underlying model that can accurately predict the behaviour of all the datasets involved. In this section, we take a step back and explore a model-influenced but largely model-independent method for using power spectrum results to enhance a global signal measurement.

Unlike power spectrum measurements, which must deal with the dual challenges of sensitivity and foreground contamination, global signal measurements are typically not lacking in sensitivity \citep{presley_et_al2015}. The necessary signal-to-noise can be achieved after $\sim\!100\,\textrm{hours}$ of high quality observations with a single dipole, whereas power spectrum measurements require roughly an order of magnitude more integration time even with the largest low-frequency arrays in existence. The chief obstacle to a successful measurement of the global signal is then foreground contamination.

Foregrounds are more difficult to deal with in global signal experiments than they are for power spectrum measurements. This is simply because fewer numbers are measured. In particular, if the global signal spectrum is measured at $N_\textrm{chan}$ different spectral channels, it is mathematically impossible to simultaneously constrain the $2N_\textrm{chan}$ degrees of freedom from $N_\textrm{chan}$ channels of the cosmological signal and the $N_\textrm{chan}$ channels of the foreground signal. One way to deal with this problem is to increase the number of measurements. For example, angular information can be recorded, increasing the number of independent measurements to $N_\textrm{chan} N_\textrm{pix}$, where $N_\textrm{pix}$ is the number of independent pixels surveyed on the sky. This extra information can then be used to aid foreground subtraction \citep{liu_et_al2013}. Alternatively, if the number of measurements is held fixed at $N_\textrm{chan}$, one can decrease the number of parameters that one is trying to constrain. In one extreme, for example, if the foreground spectrum is known perfectly, it can simply be subtracted directly from the measured spectrum, leaving only the cosmological contribution to the signal at $N_\textrm{chan}$ different frequencies. More realistically, the foregrounds are not known perfectly, and assumptions need to be made regarding the possible forms of the foreground and cosmological signal contributions. In practice, these assumptions are often encapsulated by the specific parametric forms that are chosen to fit the data.

The success of a global signal experiment thus depends on our ability to compactly describe the foregrounds and the cosmological signal in as few parameters as possible \citep{switzer_and_liu2014}. The parameterization of foreground spectra has already been considered elsewhere in the literature, with most treatments fitting out foregrounds using smooth functions like low-order polynomials. As with the previous section, we will continue to parameterize the foregrounds as a series of low-order orthogonal Legendre polynomials in $\ln \nu$. Our goal is instead to tackle the problem of how one might parameterize the cosmological signal. Two popular parameterizations of the global signal---the turning point model and the tanh model---have been recently compared in \citet{harker_et_al2015}. Here we propose a third.

Suppose one has made a successful detection of the $21\,\textrm{cm}$ power spectrum. Further suppose that using these data (possibly along with other datasets), one has inferred a set of best-fit values for a set of astrophysical and cosmological parameters (such as those shown in Table \ref{tab:Params}). These best-fit values can then be used to predict a fiducial global signal $T_{21}^\textrm{fid} (\nu)$ by running reionization simulations such as {\tt 21cmFAST}. This fiducial global signal can then be treated as first guess of the true global signal, and a global signal measurement can be used to measure \emph{deviations} from this fiducial signal. Note that in redefining the goals of a global signal measurement in this way, we have not formally improved our situation. The fiducial global signal acts only as a frequency-dependent offset in our measurement, and does not reduce the number of degrees that one must fit for. To do so, we must identify the ways in which the global signal is likely to deviate, given the uncertainties in our modeling of reionization that still remain after the power spectrum measurement. This can be done using a similar approach to the one that we adopted for the ionization history in Section \ref{sec:ionHist}. We approximate the global signal as
\begin{equation}
\label{eq:LinearGlobalSig}
\overline{T}_{21} (\nu) \approx \overline{T}_{21}^\textrm{fid} (\nu) + \sum_\gamma g_\gamma (\nu) \Delta p_\gamma,
\end{equation}
where $g_\gamma(\nu)$ is the response of the global signal to a shift in the $\gamma$th parameter from its fiducial (or best-fit) value, and $\Delta p_\gamma$ is again the size of this shift. The statistically likely deviations in the global signal can then be captured by a signal covariance matrix
\begin{equation}
\label{eq:Sdef}
\mathbf{S}_{ij} \equiv \langle \overline{T}_{21} (\nu_i)  \overline{T}_{21} (\nu_j)\rangle  - \overline{T}_{21}^\textrm{fid} (\nu_i) \overline{T}_{21}^\textrm{fid} (\nu_j),
\end{equation}
where in this equation (unlike with Eq. \ref{eq:FisherDef}) the angled brackets $\langle \dots \rangle$ signify an ensemble average over variations in the parameters, and thus $\langle \overline{T}_{21} (\nu) \rangle = \overline{T}_{21}^\textrm{fid} (\nu)$ by definition. Inserting Eq. \eqref{eq:LinearGlobalSig} into the definition of $\mathbf{S}$ then gives
\begin{equation}
\label{eq:GCG}
\mathbf{S} = \mathbf{G} \mathbf{C} \mathbf{G}^t,
\end{equation}
where $\mathbf{G}_{i\alpha} \equiv g_\alpha (\nu_i)$ and $\mathbf{C}$ is the covariance matrix of the parameters. In other words, interpreting the parameter deviations in Eq. \eqref{eq:LinearGlobalSig} as being due to uncertainties in the best-fit parameters from the power spectrum measurement, we have $\mathbf{C}_{\alpha \beta} \equiv \langle \Delta p_\alpha  \Delta p_\beta \rangle$.

\begin{figure}
\centering
\includegraphics[width=0.5\textwidth]{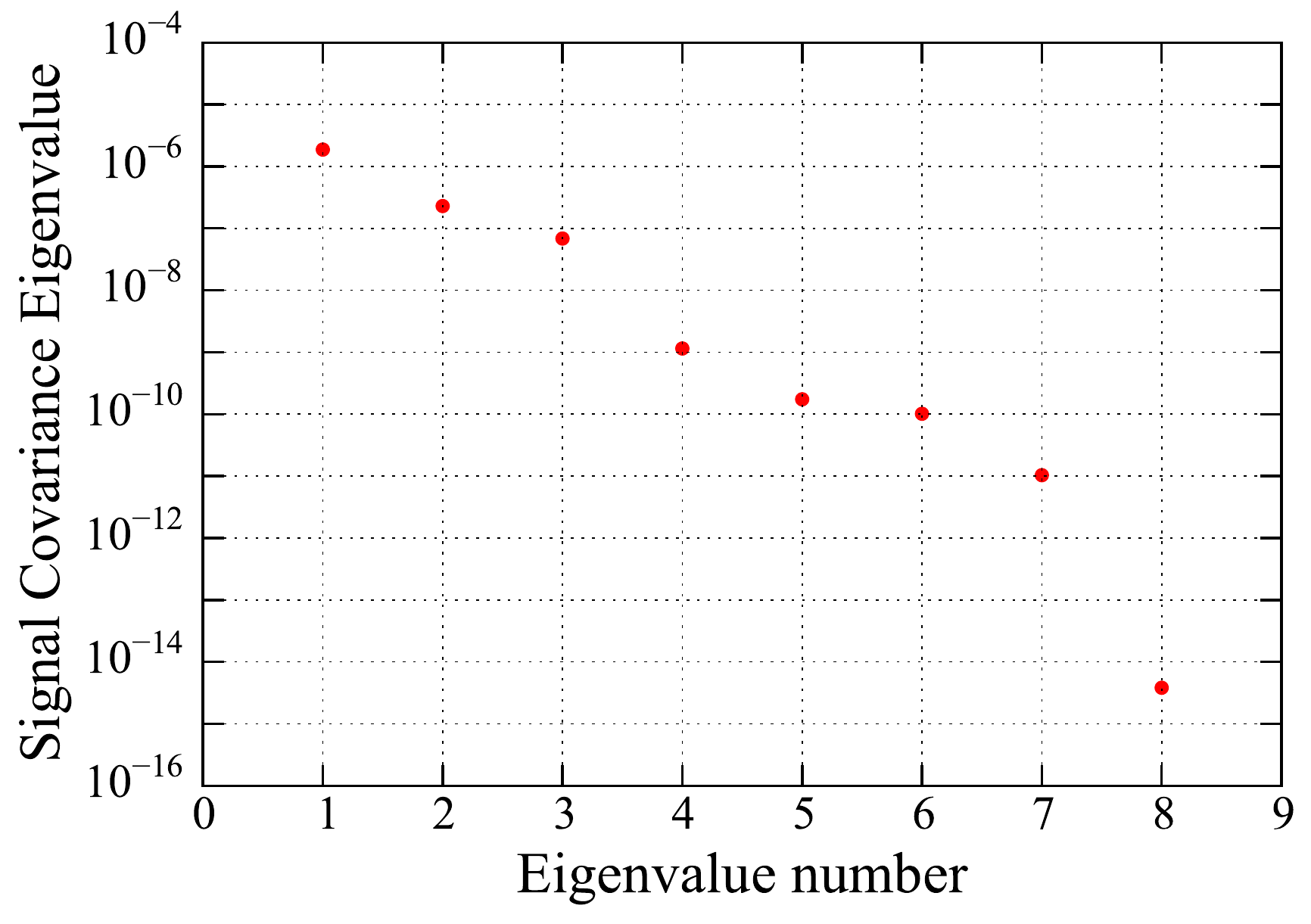}
\caption{The first few eigenvalues of $\mathbf{S}$ (Eqs. \ref{eq:Sdef} and \ref{eq:GCG}), assuming a covariance matrix $\mathbf{C}$ of astrophysical and cosmological parameters from HERA. The exponential fall-off seen in this plot suggests that once the relevant astrophysical and cosmological parameters have been measured to reasonable precision by a power spectrum experiment, the allowable deviations of the global signal from a fiducial model (encoded by $\mathbf{S}$) can be captured by a small number of parameters.}
\label{fig:scree}
\end{figure}

\begin{figure*}
\centering
\includegraphics[width=1.0\textwidth]{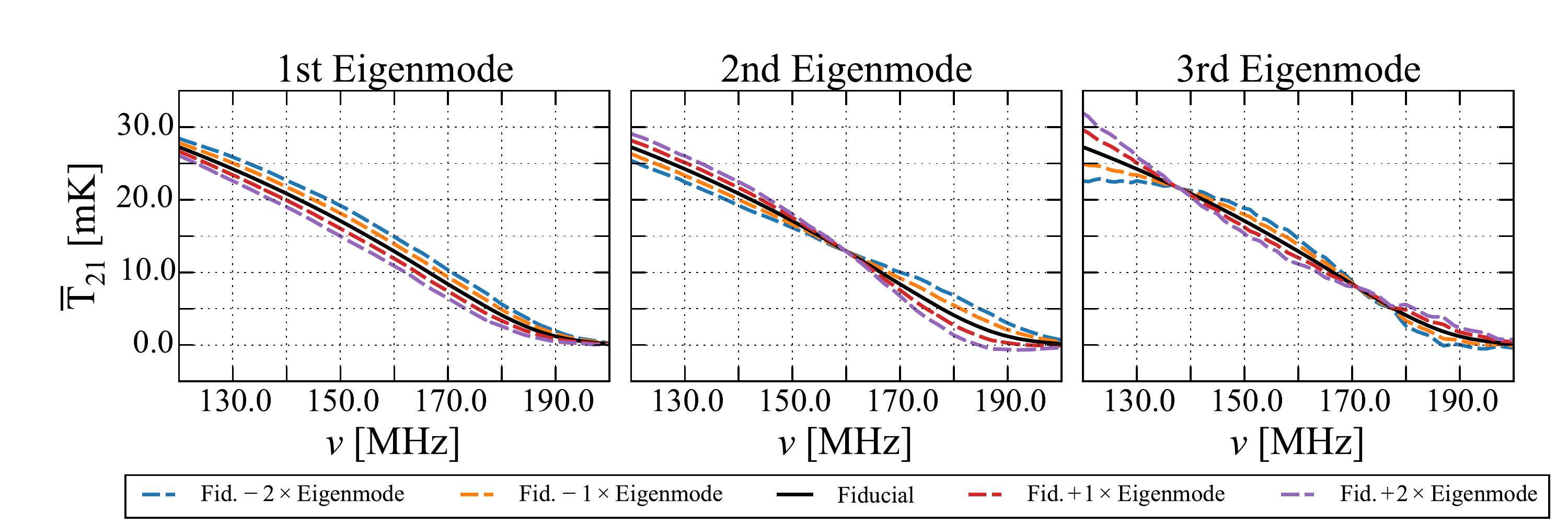}
\caption{Fiducial global signal (solid black) perturbed by the first deviation eigenmode (left plot), second eigenmode (center plot), and the third eigenmode (right plot). In each plot, both positive and negative perturbations to the fiducial model are shown. The first eigenmode is seen to mostly shift the redshift of the midpoint of reionization without altering its duration, while the second eigenmode changes the duration while keeping the midpoint constant. Higher eigenmodes add increasingly complicated features to the global signal.}
\label{fig:devEigenmodes}
\end{figure*}

To identify the spectral modes that represent the likeliest deviations from the fiducial global signal, we perform an eigenvalue decomposition of $\mathbf{S}$. If the eigenvalues are dominated by a handful of large values, then plausible forms for the global signal (in the context of the limitations imposed by the measured astrophysical and cosmological parameters and the physical constraints of reionization simulations) can be captured by just a small number of eigenmodes. We see precisely this behaviour in Fig. \ref{fig:scree}, where the eigenvalues of $\mathbf{S}$ are seen to fall off exponentially. Intuition for the first few eigenmodes can be obtained by examining Fig. \ref{fig:devEigenmodes}, where we show examples of how each eigenmode perturbs the fiducial global signal. The first eigenmode is seen to mostly control the redshift of the midpoint of reionization without affecting its duration. The second eigenmode affects the duration without affecting the midpoint. The third eigenmode accounts for higher order curvature effects in the global signal history. Taken together, these results suggest that even though the details of reionization depend on astrophysical and cosmological parameters in complicated ways, uncertainties in these model parameters can only induce rather specific types of changes in the global signal. Thus, an economical model (in the sense of having few parameters) for the cosmological global signal is given by
\begin{equation}
\label{eq:devEigenExpansion}
\overline{T}_{21} (\nu) \approx \overline{T}_{21}^\textrm{fid} (\nu) + \sum_i^{N_d} b_i d_i (\nu),
\end{equation}
where $d_i (\nu)$ is the $i$th eigenmode of $\mathbf{S}$, and it is assumed that $N_d$ such ``deviation eigenmodes" are sufficient to capture deviations from the fiducial model in the measurement. The coefficients $\{b_i \}$ are the parameters to be fit in an analysis of global signal data. Once these are determined, they can reinserted into Eq. \eqref{eq:devEigenExpansion} for an estimate of the global signal.

By expressing the signal in terms of eigenmodes of $\mathbf{S}$, our method is effectively a special case of Karhunen-Lo\`{e}ve (or signal-to-noise) data compression \citep{bond1995,bunn_and_sugiyama1995,vogeley_and_szalay1996,tegmark_et_al1997}. In the Karhunen-Lo\`{e}ve method, one derives a set of generalized eigenvectors $\mathbf{v}$ that satisfy the equation $\mathbf{S} \mathbf{v} = \lambda \mathbf{N} \mathbf{v}$, where $\lambda$ is the generalized eigenvalue and $\mathbf{N}$ is the noise covariance matrix. The signal is then expressed in terms of the eigenvectors with the largest eigenvalues. In this paper we are therefore employing a special case of this, where the noise covariance matrix is taken to be the identity (i.e., we are using ``signal eigenmodes" rather than ``signal-to-noise" eigenmodes). In our computational explorations, we find that for the present application, true Karhunen-Lo\`{e}ve modes (with $\mathbf{N} \neq \mathbf{I}$) tend to be rather sensitive to the details of $\mathbf{N}$. (In other words, many of the ``signal-to-noise" eigenmodes are driven by the behaviour of the noise rather than the signal). This makes it more difficult to provide a clean interpretation of our modes (in the style of Fig. \ref{fig:devEigenmodes} and the discussion above), and complicates the discussion of the interplay between the form of the signal and the foregrounds that we present later in this section. For this paper, then, we elect to parameterize the measurements in terms of signal eigenmodes rather than signal-to-noise eigenmodes.

Under our proposed data analysis scheme, the forecasted performance of a global signal measurement will be somewhat tied to the performance of a power spectrum experiment. This is because our parameterization of the cosmological global signal requires prior information from a power spectrum measurement, entering via Eq. \eqref{eq:GCG} as the covariance matrix $\mathbf{C}$ of astrophysical and cosmological parameters. As our worked example, we will again use HERA, obtaining $\mathbf{C}$ from the Fisher forecasts of Section \ref{eq:PlainParameterConstraints} for the \emph{Planck} TT,TE,EE+lowP+lensing+ext scenario (the results for \emph{Planck} TT+lowP being qualitatively similar). Importantly, however, our global signal forecasts do not depend on the \emph{amplitude} of the parameter errors from the power spectrum measurements, only the \emph{shapes} of parameter degeneracies. To see this, notice that rescaling $\mathbf{C}$ by an overall constant results only in a corresponding rescaling of $\mathbf{S}$, which affects its eigenvalues but not its eigenvectors (i.e., the deviation eigenmodes). Such a rescaling therefore has no effect on the parameterization given by Eq. \eqref{eq:devEigenExpansion}, and once the deviation eigenmodes $\{ d_i (\nu) \}$ are determined we no longer require knowledge of our power spectrum experiment.

\begin{figure}
\centering
\includegraphics[width=0.52\textwidth]{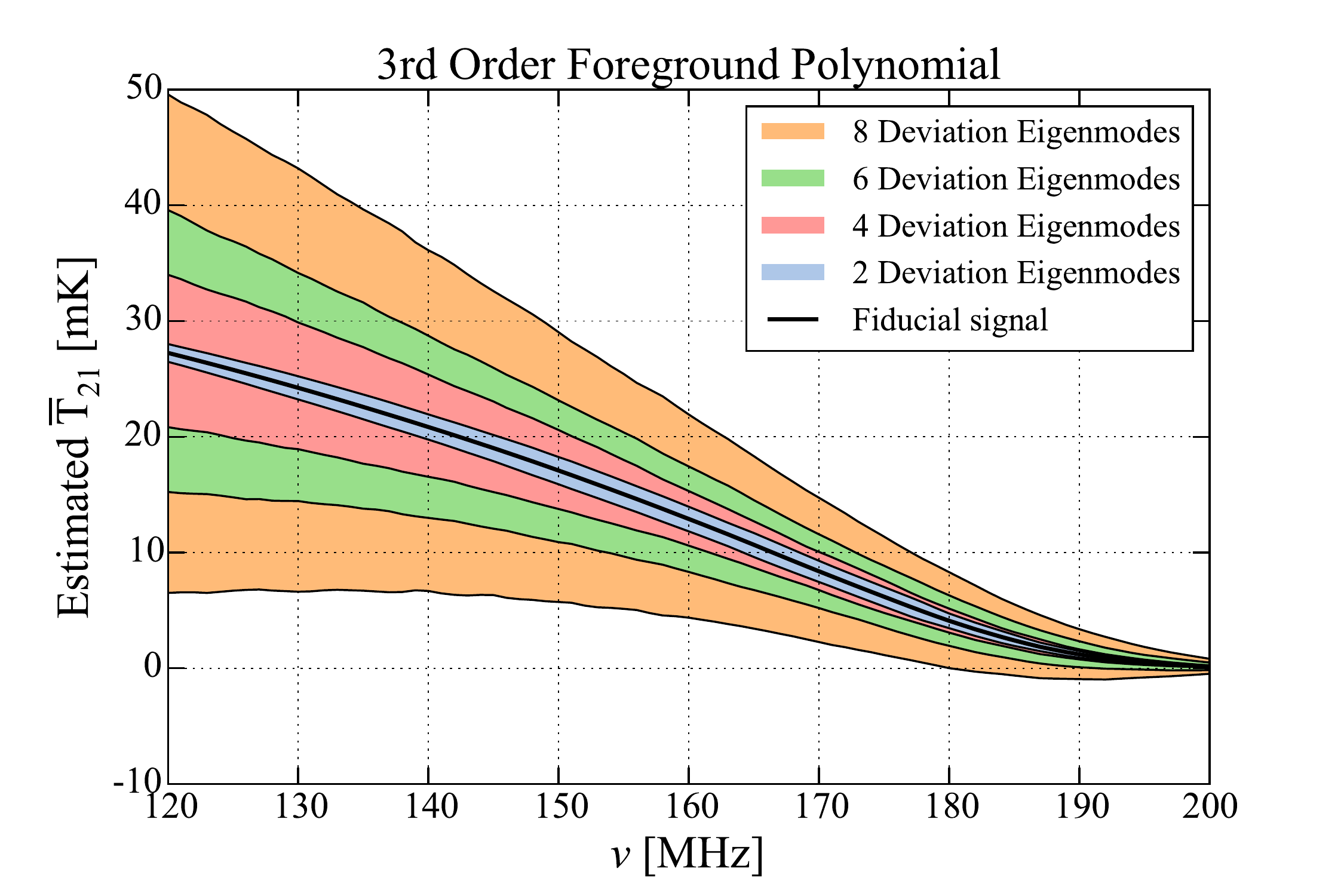}
\caption{Forecasted global signal recovery, assuming that foregrounds can be adequately described by a third order (i.e., cubic) polynomial in $\ln \nu$. The quality of the recovery depends on the number of deviation eigenmodes that are needed to obtain a good fit to the cosmological global signal. Each coloured band shows the $95\%$ confidence region of the recovered signal for a different number of deviation eigenmodes.}
\label{fig:FgPoly3ConfidenceRegion}
\end{figure}

\begin{figure}
\centering
\includegraphics[width=0.52\textwidth]{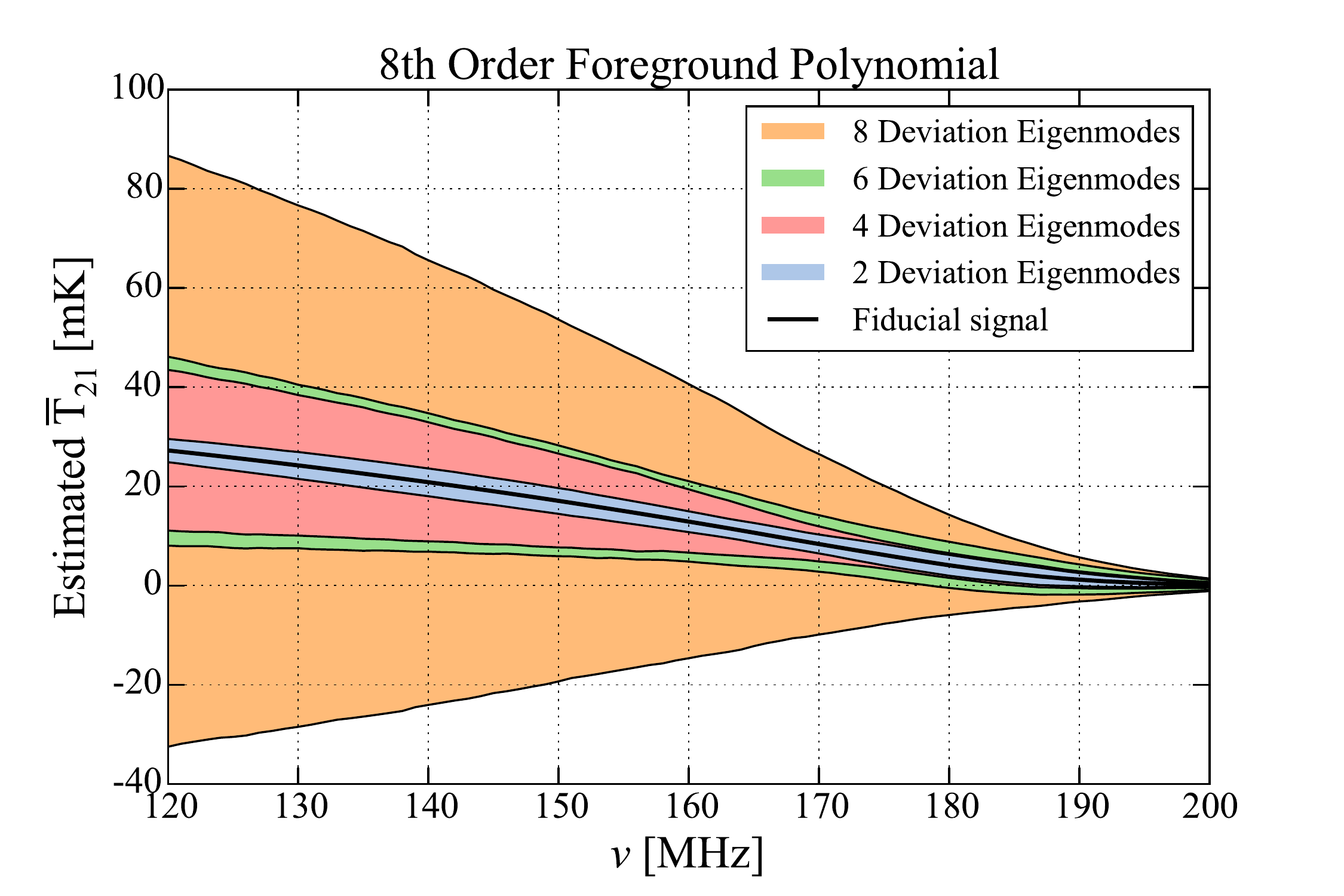}
\caption{Same as Fig. \ref{fig:FgPoly3ConfidenceRegion} but assuming that a successful foreground fit requires an eighth order polynomial in $\ln \nu$. Compared to Fig. \ref{fig:FgPoly3ConfidenceRegion}, the recovery of the global signal is clearly degraded. However, reasonably precise constraints can still be obtained if deviations from the fiducial model can be adequately described by a small number of deviation eigenmodes.}
\label{fig:FgPoly8ConfidenceRegion}
\end{figure}

With deviation eigenmodes in hand, we may push forward to a global signal forecast, again employing the Fisher matrix formalism. The procedure is essentially the same as the one employed in the previous section, except with the expansion coefficients of Eq. \eqref{eq:devEigenExpansion} taking the place of the astrophysical and cosmological parameters. The foreground-marginalized Fisher forecasts are then used to draw random realizations of $\{b_i \}$, which are in turn propagated into confidence regions for $\overline{T}_{21} (\nu)$ using Eq. \eqref{eq:devEigenExpansion}. The results are shown in Figs. \ref{fig:FgPoly3ConfidenceRegion} and \ref{fig:FgPoly8ConfidenceRegion}, assuming that foregrounds can be adequately fit using polynomials of order three and eight, respectively. The former is intended to represent the relatively optimistic assumption of an instrument with a largely achromatic beam response that does not imprint extra spectral structure onto the smooth foregrounds \citep{pritchard_and_loeb2010,morandi_and_barkana2012}, while the latter is intended to be reflective of a more pessimistic assumption where chromatic beams or other effects degrade the smoothness of the observed foregrounds \citep{vedantham_et_al2014,bernardi_et_al2015,presley_et_al2015}. In each plot, we show the $95\%$ confidence regions of the recovered global signal, with the different colours corresponding to different assumptions regarding the number of eigenmodes that are needed to adequately fit any deviations from the fiducial spectrum.

From these plots, it is clear that the global signal can be measured with high precision if very few deviation eigenmodes are needed to fit the data. For example, we see that if only two modes are necessary, either foreground scenario will yield tight constraints on the global signal. This is perhaps unsurprising, since we saw from Fig. \ref{fig:devEigenmodes} that the first two modes mostly encode information about the timing of reionization, which previous studies have suggested can be inferred reasonably well from global signal experiments \citep{pritchard_and_loeb2010,morandi_and_barkana2012,liu_et_al2013,presley_et_al2015}. As the number of necessary deviation eigenmodes increases, however, a comparison of Figs. \ref{fig:FgPoly3ConfidenceRegion} and \ref{fig:FgPoly8ConfidenceRegion} reveals that it is crucial that foreground contaminants be describable by a small number of smooth components.

In this section, we have shown how $21\,\textrm{cm}$ power spectrum measurements can be used to construct a parameterization for the cosmological global signal. The resulting parameterization is designed to economically describe likely deviations from a fiducial global signal using only a small number of parameters, thus increasing the likelihood of a detection of the global signal amidst foreground contaminants. Importantly, we note that while the parameterization is certainly influenced by the power spectrum measurements and the theoretical models used to interpret them, our scheme is formally model-independent. This is because the eigenmodes of $\mathbf{S}$ form a basis that spans the space of all possible spectra.\footnote{Note that this is true even though $\mathbf{S}$ is in general not a full-rank matrix, and will typically have the same number of non-zero eigenvalues as $\mathbf{C}$. Even without a full set of non-zero eigenvalues, the eigenvectors of $\mathbf{S}$ span the space. To see this, recall that adding the identity matrix to a rank-deficient $\mathbf{S}$ has no effect on its eigenvectors, but will make all the eigenvalues non-zero.} As a result, even arbitrary deviations from the fiducial model can in principle be captured in our analysis, simply by increasing the number of deviation eigenmodes that one fits for. Our method is thus model-independent, although in practice the precision of the recovered global signal will degrade with the number of modes. One should therefore use as few components as are necessary, using model-selection algorithms to select the optimal number of deviation eigenmodes that are demanded by the data, analogous to the way in which \citet{harker2015} selected the number of foreground components. In fact, the ideal treatment would be to perform a joint model-selection exercise that simultaneously determines the number of foreground and cosmological signal parameters. In any case, by using power spectrum results to inform our parameterization, we increase the likelihood of being able to use a small number of components while preserving formal model independence.

Although we saw in Section \ref{sec:BruteJointFit} that global signal measurements may not inform parameter constraints, they are still valuable in providing a cross-check on the results of a power spectrum measurement. The most fruitful 
way to enact such a cross-check will be to check for consistency in the timing of reionization, since Fig. \ref{fig:devEigenmodes} shows that global signal measurements will be particularly sensitive to timing. Moreover, \citet{liu_et_al2015} recently showed that as a direct measurement of the brightness temperature field, the global signal can provide a more model-independent way to remove reionization as a nuisance effect in CMB studies. Finally, our treatment of the global signal in this paper pertains only to those experiments targeting the reionization epoch exclusively. At higher redshifts the global signal takes a more complicated form, and experiments will be able to place precise constraints on the astrophysics of the IGM \citep{harker_et_al2012,mirocha_et_al2013,mirocha_et_al2015}.

\section{Conclusions}
\label{sec:Conclusions}

In this paper, we have addressed a number of outstanding issues in forecasts for upcoming $21\,\textrm{cm}$ power spectrum and global signal measurements. First, we provided updated forecasts for $21\,\textrm{cm}$ power spectrum measurements seeking to constrain astrophysical and cosmological parameters. Our forecasts are based on two sets of fiducial parameters from the \emph{Planck} satellite. One set is based on the \emph{Planck's} TT+lowP dataset and features a relatively high optical depth $\tau$, while the other is based on \emph{Planck's} TT,TE,EE+lowP+lensing+ext dataset and features a relatively low $\tau$. Using a Fisher matrix formalism, we find that the projected parameter constraints from a power spectrum measurement are better for the TT,TE,EE+lowP+lensing+ext dataset than the TT+lowP dataset. The lower $\tau$ of the latter dataset implies a lower redshift of reionization, which is favourable to $21\,\textrm{cm}$ experiments. This is because lower redshifts translate into higher frequencies for $21\,\textrm{cm}$ observations, where the foregrounds are dimmer and instrumental noise is lower. Additionally, the cosmological constraints from the TT,TE,EE+lowP+lensing+ext dataset alone (i.e., without $21\,\textrm{cm}$ information) are tighter than those from TT+lowP. This also contributes to better parameter constraints because instruments like HERA are sensitive enough to make cosmological parameter uncertainties non-negligible in one's data analysis.

Turning to angle-averaged quantities, we have provided the first forecasts of HERA's ability to constrain the ionization history. These forecasts compare favourably to those provided by non-$21\,\textrm{cm}$ probes, significantly narrowing the set of plausible histories even if low levels of modeling errors are present in semi-analytic simulations. We also show forecasts of astrophysical and cosmological parameter constraints from joint fits over both power spectrum and global signal measurements. Unfortunately, we find that a global signal measurement limited to the likely redshift range of reionization (say, between $z\sim 6$ and $11$) will not add much to the model constraints already provided by power spectrum measurements. This is because the global signal is relatively featureless during reionization, making parameter variations difficult to pick out from shifts in foreground parameters. As suggested in \citet{mirocha_et_al2015}, direct constraints on model parameters from global signal experiments will likely require broadband instruments that reach beyond reionization to higher redshifts.

Importantly, however, we note that it is still important to have global signal experiments that target reionization, because our description of the $21\,\textrm{cm}$ brightness temperature field will be incomplete without them. Unlike the CMB, whose sky-averaged signal follows a simple blackbody spectrum at leading order, the mean $21\,\textrm{cm}$ signal (i.e., the global signal) has a non-trivial frequency dependence due to heating and ionization processes in the IGM. A complete description of the $21\,\textrm{cm}$ brightness thus requires not only a quantification of the spatial fluctuations (via the power spectrum, for example), but also the global signal. The global signal additionally provides a good cross-check on power spectrum results, and can enhance CMB constraints in a more model-independent way \citep{liu_et_al2015}. In this paper, we have proposed a method by which the results from a power spectrum measurement can be used to aid a detection of the global signal. Power spectrum measurements are used to measure the underlying parameters of reionization simulations. The reionization simulations are then used to predict plausible global signal histories, with the covariance of the measured parameters translating into a covariance of global signal histories. An eigenvalue decomposition of the latter covariance then provides a basis of ``deviation" eigenmodes whose amplitudes can be constrained by global signal experiments. Typically, only a small number of these eigenmodes will represent deviations from fiducial global signal histories that are consistent with both the physical constraints imposed by the simulations and the power spectrum measurements. Thus, if deviations from fiducial models are relatively small, only a few modes will be necessary to fit the global signal data, enabling a high signal-to-noise measurement. On the other hand, if deviations are large, more modes will be required and the measurement will be degraded. However, because the deviation eigenmodes collectively span the space of all possible global signal histories, our scheme will still allow accurate (if not precise) measurements of the global signal. In other words, while our eigenmode parameterization of the global signal is model-\emph{influenced}, it is not formally model-\emph{dependent}.

This paper reaffirms the promise of $21\,\textrm{cm}$ cosmology, showing that even when cosmological parameter uncertainties are included in one's forecasts, the resulting constraints on astrophysical parameters from power spectrum measurements are still excellent. Moreover, these constraints can then be used to train a spectral basis set for fitting global signal data, boosting the potential for global signal experiments to provide an independent window into the high redshift universe. With improved instruments like HERA being constructed, our forecasts confirm that $21\,\textrm{cm}$ cosmology has the potential to significantly improve our understanding of astrophysics and cosmology.

\section*{Acknowledgements}

It is a pleasure to acknowledge Jonathan Pober for helpful discussions, and Bradley Greig and Andrei Mesinger for helpful discussions and for providing the combined ionization history constraint data in Fig. \ref{fig:ionHist}. This research was completed as part of the University of California Cosmic Dawn Initiative. AL and ARP acknowledge support from the University of California Office of the President Multicampus Research Programs and Initiatives through award MR-15-328388, as well as from NSF CAREER award No. 1352519, NSF AST grant No.1129258, and NSF AST grant No. 1440343. AL acknowledges support for this work by NASA through Hubble Fellowship grant \#HST-HF2-51363.001-A awarded by the Space Telescope Science Institute, which is operated by the Association of Universities for Research in Astronomy, Inc., for NASA, under contract NAS5-26555.




\bibliographystyle{mnras}
\bibliography{refs} 



%
%
%


\bsp	
\label{lastpage}
\end{document}